# Electronic Collective Variables for Chemical Reaction Sampling

*YaoKun Lei[1], Isaac Yang[1,*]*

[1] Institute of Systems and Physical Biology, Shenzhen Bay Laboratory, Shenzhen 518132, China

* To whom correspondence should be addressed: yangyi@szbl.ac.cn

**Abstract**

Chemical reaction sampling critically depends on collective variables (CVs) that capture the slow degrees of freedom governing reactive transformations. However, existing reaction CVs are often defined in geometric space or learned in a system-specific manner, which limits their transferability and leaves open the more fundamental question of how reaction progress should be represented. From a physical perspective, chemical reactions are defined by electron redistribution. Here, we introduce a charge-space electronic collective variable that describes the electronic component of reaction progress in a common linear form based on atomic charges. To enable its use in enhanced sampling, atomic charges and the corresponding CV gradients are provided by a neural-network model trained on QM/MM data within an iterative sampling-training workflow. Across multiple reactions in aqueous and enzymatic environments, we show that this electronic CV can be constructed in a common charge-space form, with the corresponding coefficients assigned in a simple manner from charge differences between relevant states. Our simulations further show that reaction progress generally involves coupled electronic and conformational components, and that the same framework can also be extended to restrain side reactions. These findings support charge-based electronic CVs as a physically motivated framework for describing the electronic component of chemical reaction progress with reduced reliance on handcrafted geometric descriptors.

# 1. Introduction

Chemical reactions are central to catalysis, enzymatic function, and synthetic chemistry, and their theoretical description depends on adequate sampling of reactive configurations along the reaction pathway. Such sampling is important not only for mechanistic analysis and free-energy calculations but also for constructing high-quality datasets in the data-driven era, including those used to train neural-network potentials[1, 2], predict molecular properties[3, 4], and support molecular generation and design[5]. In practice, however, reactive events are rare on conventional molecular dynamics time scales, particularly near transition regions where subtle structural variations can lead to large energetic changes. Enhanced sampling methods provide an effective route to overcome this limitation[6-8], but their success depends critically on the choice of collective variables (CVs) that capture the slow degrees of freedom governing reaction progress. Developing generally applicable CVs for chemical reactions therefore remains a major challenge in molecular simulation.

Most existing CVs for reaction sampling are defined in geometric space[9], typically in terms of distances, angles, or dihedral angles associated with bond breaking, bond formation, and conformational rearrangement. Although such descriptors can be effective for specific systems, their construction often requires substantial domain knowledge and careful system-dependent tuning. Recently, machine-learning-based approaches have been introduced to learn optimized CVs directly from simulation data[10-12], including methods based on committor-like objectives[13-20] and transfer-operator formulations[21-23]. Although these approaches provide an automated route for CV construction, their generality mainly lies in the optimization objective rather than in the physical representation of the CV itself. Much less attention has been paid to the more fundamental question of how reaction progress should be represented and in which descriptor space a more transferable reaction CV should be constructed. This suggests that what is still missing is a common physically grounded representation for the electronic component of reaction progress, upon which more transferable reaction CVs may be constructed.

From a physical perspective, the defining feature of a chemical reaction is electron redistribution. A central difficulty in constructing transferable reaction CVs is that different reactions exhibit markedly different patterns of electron redistribution in the atom-type-dependent Cartesian space, while the underlying electron redistribution within each reaction is itself highly nonlinear and nonlocal. Because of this complexity, it is often difficult to represent reaction progress in geometric space using a common functional form that transfers naturally across systems. Motivated by this difficulty, we seek a more suitable representation of reaction progress in electronic space. Because electronic descriptors are intended to characterize specific patterns of electron distribution, the desired CV can in principle be related to them through a relatively simple mathematical function. This provides a natural basis for a common CV-design strategy with simple coefficient assignment. In this work, we use atomic charges as simple coarse-grained descriptors of electron distribution. In this sense, they provide a zeroth-order representation of the dominant electronic redistribution relevant to reaction progress. We then construct the electronic CV in a common linear form to describe the electron-redistribution component of different reactions.

Using this framework, we investigate several representative reactions in both aqueous and enzymatic environments. We show that, across all tested systems, the electronic CV can be constructed in a common charge-space form as a linear combination of atomic charges, with the corresponding coefficients assigned in a simple manner from charge differences between relevant states; the same framework can also be extended to restrain side reactions. More importantly, our simulations show that reaction progress generally involves two coupled components: a conformational component that drives the structural adjustment needed to reach reactive configurations and an electronic component that captures the redistribution of electrons. These results support charge-based electronic CVs as a physically motivated framework for resolving the electronic component of reaction progress with reduced reliance on handcrafted geometric descriptors.

## 2. Method
### 2.1 CV Design

Chemical reaction progress generally involves two coupled components: conformational adjustment and electron redistribution. Accordingly, an effective CV description of reaction progress should account for both components. Motivated by this observation, we construct the reaction CV from an electronic component and, when necessary, a complementary conformational component. The electronic component is intended to capture the redistribution of electrons associated with chemical transformation, whereas the conformational component facilitates the structural rearrangements required to access reactive configurations.

To construct a common descriptor space for the electronic component of reaction progress, we use atomic charges as simple coarse-grained descriptors of electron redistribution. In this sense, atomic charges provide a zeroth-order approximation to the underlying electron distribution, while higher-order electronic descriptors may be incorporated in future extensions. We therefore define an electronic collective variable in charge space as

$$CV_{elec} = f\left(q_1, \ldots, q_{N_{QM}}; \mathbf{T_{MM}}\right), \quad (1)$$

where $q_i$ denotes the quantum atomic charge of the i-th QM atom, $N_{QM}$ is the number of QM atoms in the QM/MM region, and the electrostatic environment generated by the surrounding MM atoms is represented by $\mathbf{T_{MM}}$. In the present work, the reference charges are taken from DFTB3[24]/MM calculations. Within this charge-space representation, we use the following common linear form as a first-order approximation to the dominant electronic change along the reaction process:

$$CV_{elec-ref} = \sum_{i=1}^{N_{QM}} c_i q_i, \quad (2)$$

where $c_i$ is the coefficient associated with atom i. In the simplest construction, the coefficient is assigned from the charge change between the reactant and product states according to

$$\delta q_i = q_i(product) - q_i(reactant), \tag{3}$$

$$c_i = \begin{cases} sign(\delta q_i) & |\delta q_i| > \epsilon \\ 0 & |\delta q_i| \leq \epsilon \end{cases}, \tag{4}$$

where $q_i(product)$ and $q_i(reactant)$ denote the reference charges of atom i in the reactant and product states, respectively, and $\epsilon$ is a threshold used to exclude atoms whose charge changes are too small to contribute meaningfully to the electronic component of reaction progress. This choice ensures that the electronic CV follows the dominant direction of electron redistribution while suppressing minor fluctuations. In more challenging cases, the coefficients can instead be determined by linear discriminant analysis (LDA, see section 4.2 in Supporting Information), which projects the charge space onto a one-dimensional coordinate that best distinguishes the relevant states or basins. Overall, this construction requires only representative structures from the two end-state basins and therefore makes the electronic CV straightforward to define while substantially reducing the need for handcrafted, system-specific design.

Since analytical gradients of the reference quantum atomic charges are not readily available, we instead train a neural-network model $Q_\theta$ to fit the charges and obtain the gradients required for biased simulations:

$$\hat{q}_i = Q_\theta(\mathbf{Z_{QM}}, \mathbf{R_{QM}}, \mathbf{T_{MM}}), \tag{5}$$

where $\hat{q}_i$ denotes the predicted charge, $\mathbf{Z_{QM}}$ and $\mathbf{R_{QM}}$ denote the nuclear charges and Cartesian coordinates of the QM atoms, respectively, and $\mathbf{T_{MM}}$ denotes the surrounding MM electrostatic environment. The predicted electronic CV $\text{CV}_{\text{elec}}$ and its gradient $\nabla_R \text{CV}_{\text{elec}}$ are then evaluated as

$$\text{CV}_{\text{elec}} = \sum_{i=1}^{N_{\text{QM}}} c_i \hat{q}_i, \tag{6}$$

$$\nabla_R \text{CV}_{\text{elec}} = \sum_{i=1}^{N_{\text{QM}}} c_i \nabla_R \hat{q}_i. \tag{7}$$

Because the initial training data typically cover only the basin region, enhanced sampling and model retraining are carried out iteratively. The neural-network fitting of charges and the iterative workflow are illustrated in Figure 1. In practice, two or three iterations are usually sufficient because the accuracy required for CV construction is less stringent than that required for a production-quality energy model.

The electronic CV alone does not necessarily accelerate the conformational adjustment required for reaction. We therefore consider two complementary strategies to enhance the conformational component. First, integrated tempering sampling (ITS)[25] can be used to increase the probability of high-energy conformations without requiring detailed domain knowledge. However, increasing the accessible energy range does not necessarily accelerate a specific conformational change efficiently because the accessible configurational space becomes much larger in the high-energy region. A second strategy is to explicitly define a general conformational CV as

$$CV_{\text{conf}} = g(r_1, r_2, \ldots, r_M). \tag{8}$$

where $r_j$ denotes the selected geometrical descriptor. In the present work, however, we find that a simple linear combination of distances associated with forming $d_{bonding}$ and breaking bonds $d_{breaking}$ is already effective for the tested systems,

$$CV_{conf} = d_{bonding} - d_{breaking}. \tag{9}$$

The resulting biasing variable used in simulation is then written as

$$CV = c_{conf} * CV_{conf} + c_{elec} * CV_{elec}, \tag{10}$$

where $c_{conf}, c_{elec}$ are linear coefficients.

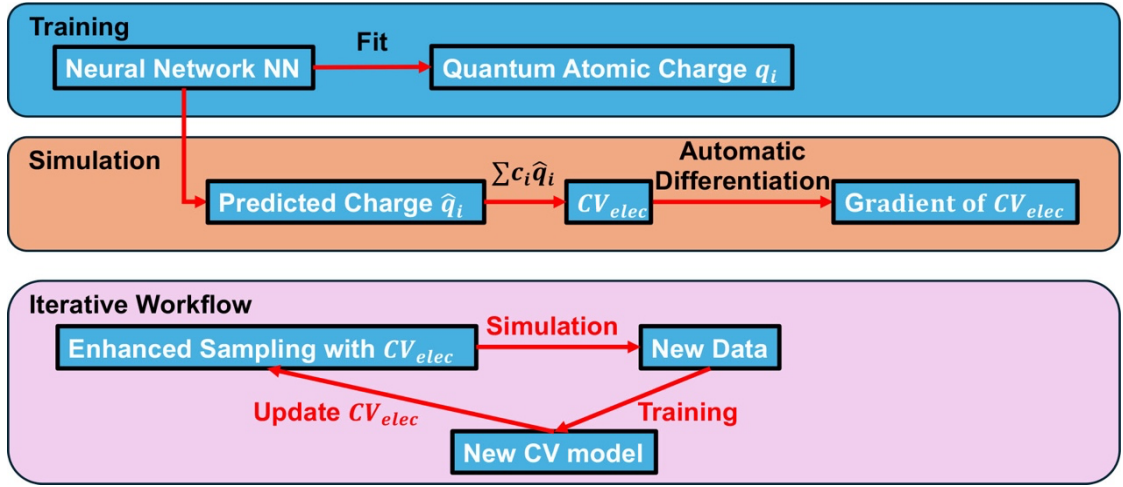

Figure 1. Workflow of neural-network-assisted electronic CV evaluation. Top: a neural network is trained to fit atomic charges from reference quantum calculations. Middle: the trained model is used to infer atomic charges and evaluate the electronic CV and its gradients for enhanced sampling simulations. Bottom: iterative sampling-training workflow. An initial dataset is first used to train the CV model. The optimized model is then used in enhanced sampling simulations to explore the conformational space and generate additional configurations. The newly sampled data are subsequently added to the training set, and the workflow is iterated.

2.2 ML/MM Network

To evaluate the charge-based electronic CV in QM/MM simulations, we employed a previously developed ML/MM model[26, 27]. to predict the QM/MM energy and quantum atomic charges. In the present work, this model serves as a surrogate model for evaluating the atomic charges and the corresponding gradients of the electronic CV required for enhanced sampling. Here, we summarize only the aspects most relevant to the present study. In this work, the reference atomic charges are taken from DFTB3/MM calculations. As newly sampled configurations expand into regions that are not sufficiently represented in the current training set, the model is retrained iteratively to maintain reliable charge predictions and electronic-CV gradients during sampling.

The model is based on an equivariant message-passing framework, in which each QM atom is represented by scalar and vector features, $s_i^t \in \mathbb{R}^F$ and $\vec{v}_i^t \in \mathbb{R}^{F \times 3}$, that are updated through a stack of interaction layers. A key aspect of the present framework is that the effect of the surrounding MM environment on the reactive QM region is not represented through explicit high-dimensional MM coordinates, but is instead coarse-grained into an external-potential descriptor $\{T^t(\vec{R}_{QM})\}$, which is coupled to the QM atomic features throughout the interaction layers. This treatment is physically motivated by the fact that electron redistribution in the reactive region is strongly influenced by the external electrostatic potential generated by the environment. The t-th joint update of the QM features and the external descriptor is written as

$$s_i^{t+1}, \vec{v}_i^{t+1}, \{T^{t+1}(\vec{R}_{QM})\} = f_{int}^t(\{s_j^t, \vec{v}_j^t, \vec{R}_{ij} | j \in QM\}, \{T^t(\vec{R}_{QM})\}), \quad (11)$$

where $\vec{R}_{ij}$ is a relative vector between the i-th and j-th QM atoms. The external descriptor is expressed as

$$\{T^t(\vec{R}_{QM})\} = \{T_i^{(k),t}(\vec{R}_i) | k \in \mathbb{N}, i \in QM\}. \quad (12)$$

and is decomposed into MM and QM contributions,

$$T_i^{(k),t}(\vec{R}_i) = T_{MM,i}^{(k)}(\vec{R}_i) + T_{QM,i}^{(k),t}(\vec{R}_i). \quad (13)$$

Here, $T_{MM,i}^{(k)}(\vec{R}_i)$ is defined from the kth-order Taylor expansion of the electrostatic potential generated by the surrounding MM atoms

$$\{T_{MM}(\vec{R}_{QM})\} = \{T_{MM,i}^{(k)}(\vec{R}_i) | i \in QM, k \in \mathbb{N}\}, \quad (14)$$

$$T_{MM,i}^{(k)}(\vec{R}_i) = \sum_{j \in MM} q_j^{MM} T^{(k)}(\vec{R}_{ij}), \quad (15)$$

$$\left(T^{(k)}(\vec{R}_{ij})\right)_{\alpha\beta\ldots\gamma} = \nabla_\alpha \nabla_\beta \ldots \nabla_\gamma \left(\frac{1}{\|\vec{R}_{ij}\|}\right), \quad (16)$$

where $q_j^{MM}$ is the atomic charge of the j-th MM atom in the classical force field.

Analogous to $T_{MM,i}^{(k)}(\vec{R}_i)$, the QM contribution $T_{QM,i}^{(k),t}(\vec{R}_i)$, represents the long-range response between QM atoms and is updated through the interaction layers. In this way, environmental effects are incorporated through a compact and physically meaningful external-potential representation acting on the reactive QM region, rather than through explicit high-dimensional MM coordinates. This representation is central to the present framework, because it allows changes in the surrounding environment to be encoded within the same descriptor space used for charge-based CV evaluation. The detailed update expressions for the external descriptor and the interaction layers are given in our previous work[26, 27].

After the final interaction layer, the scalar features are passed to a multilayer perceptron to predict atomic energies and atomic charges. The predicted atomic charges are then used to evaluate the electronic CV and its gradients in biased simulations. To enforce the correct total charge of the QM region, the difference between the predicted net charge and the target net charge is uniformly redistributed over the QM atoms. The final energy consists of two parts: the sum of the predicted atomic energies and a Coulomb term evaluated from the inferred QM charges. In practice, the inclusion of energies and forces in training provides additional information about the local variation of the potential-energy surface, which helps constrain the learned representation and yields more reliable CV gradients during sampling. More details of the model architecture and implementation can be found in previous work[26, 27].

## 3. Simulation and Training Setup
### 3.1 Simulation Setup

System building was carried out in the QM/MM module of CHARMM-GUI[28, 29]. QM/MM simulations were carried out by combining DFTB3[24] in DFTB+[30, 31] with the GPU-native molecular dynamics package SPONGE[32] through a Python interface. 3ob-3-1 parameters[33] were used in DFTB3/MM. As illustrated in Figure 1, the simulation and model-training procedure was performed iteratively because the surrogate model used for charge-based CV evaluation must remain reliable as sampling expands beyond the reactive region covered by the initial training set.

In the first iteration, an initial training dataset containing 20,000 structures was collected in the reactant basin from a 2 ns enhanced sampling simulation using the DFTB3/MM potential together with ITS. The resulting dataset was then used to train the ML/MM neural-network model for 600 epochs with the settings described below.

In the second iteration, metadynamics[34] simulations were carried out using the electronic and conformational CVs to further explore the reaction pathway. When newly sampled configurations entered regions that were insufficiently represented in the existing training set, the prediction accuracy of the charge model deteriorated. The newly sampled configurations were therefore appended to the initial training dataset, and the ML/MM model was retrained with the augmented dataset.

From the third iteration onward, the same workflow as in the second iteration was repeated until the training set was considered sufficient to support stable charge-based CV evaluation in the relevant reactive region. In practice, this procedure typically required only two to three iterations, and starting from an initial dataset of 20,000 structures, the final training set usually involved the addition of only several hundred to fewer than 2,000 new configurations (See Tables S1, S3, S4, and S6). A production metadynamics simulation was then performed for 4 ns using the DFTB3/MM potential to estimate the free-energy profile. More details about system building, simulation workflow, and the coefficients of the electronic CV are provided in the Supporting Information.

### 3.2. Model Structure

Model hyperparameters were set as follows. The number of interaction layers was 2. The feature dimension and the number of Gaussian basis functions were both set to 64. The cutoff radius for neighborhood message calculation was 3 Å.

### 3.3 Training Setup

In addition to atomic charges, energies and forces were also included in the regression to improve model performance. This is not only to improve the overall surrogate quality, but also because direct reference gradients of the electronic CV are not readily available. By including energies and forces, the model is provided with additional information about the variation of the potential-energy surface, which helps constrain the learned charge representation and yields more reliable CV gradients during biased sampling. Therefore, the loss function $l_\text{tot}$ comprises the sum of the mean square losses of energy $l_\text{ene}$, charges of QM atoms $l_q$, and forces acting on both QM $l_\text{f,qm}$ and MM atoms $l_\text{f,mm}$.

$$l_\text{tot} = w_1 l_\text{ene} + w_2 l_\text{f,qm} + w_3 l_q + w_4 l_\text{f,mm}. \qquad (25)$$

Here, $\{w_i | i = 1 - 4\}$ are the corresponding weights. MM atoms within 10 Å from the QM region are used for training the forces. All these labels were generated by DFTB3/MM calculations. Although the model is trained with standard regression losses, the charge-based electronic CV does not require fully quantitative prediction at every configuration; rather, it only needs to provide a reasonable description of the reaction path and the associated charge redistribution. At each iteration, 600 epochs were used for training the model. More details can be found in SI.

## 4. Results

We first applied the proposed electronic CV to the Michael addition of 6'-deoxychalcone (Figure 2a) in aqueous solution and in the active site of chalcone isomerase (CHI) to illustrate why effective reaction sampling must distinguish between conformational and electronic components of reaction progress. This system is particularly suitable because the reaction proceeds through two coupled stages: a conformational adjustment that brings the attacking oxygen atom close to the electrophilic carbon, followed by electron redistribution associated with bond reorganization (Figure 2b). Accordingly, it provides a representative example for examining how geometric and electronic CVs describe different components of the same reactive transformation.

To highlight the difficulty of defining a transferable reaction CV in geometric space, we first constructed a geometry-based CV using substantial chemical intuition, as shown in Figure 2c. The bonding distance was used to describe the structural approach required for ring closure, whereas angular and dihedral terms were introduced to account for ring contraction and conjugation effects. Despite this tailored construction, the resulting geometrical CV still sampled the reactive transformation inefficiently as shown in the middle panel of Figure 3a, indicating that the electronic component of the reaction is not easily represented in a compact form in geometric space.

As a comparison, we then performed simulations using only the electronic CV. The linear coefficients of the electronic CV were assigned according to the sign of the atomic-charge changes from reactant to product (See Table S2). Figure 3b shows that the electronic CV remains nearly unchanged when the reacting atoms are still far apart, indicating that it primarily captures the electron-redistribution component of the reaction rather than the structural approach required to initiate it. Consequently, the overall transformation frequency is still limited by the efficiency of conformational adjustment as shown in the middle panel of Figure 3a. This result clearly shows that neither a purely geometric nor a purely electronic variable is sufficient by itself to describe the full reaction progress efficiently.

To enhance the conformational component in this aqueous system, we combined the electronic CV with ITS and performed Meta-ITS[35] simulations. With conformational acceleration provided by ITS, the reactive transformation became markedly more frequent as shown in the bottom panel of Figure 3a, demonstrating that efficient sampling is achieved when the electronic and conformational components are promoted simultaneously. The resulting free-energy surfaces in Figure 3b-c further clarify the distinct roles of the two coordinates. The bond-forming distance mainly reflects the conformational approach and does not fully resolve the free-energy difference between reactant and product states. Although the electronic CV more directly captures the free-energy change associated with the reaction process, it also compresses configurations with O1-C17 > 3 Å into a very narrow range of values. Consequently, the entropic effect associated with these conformational degrees of freedom is largely absent from the electronic coordinate. These results show that the electronic CV is most effective when used as a descriptor of the electronic component of reaction progress and combined with a complementary strategy that enhances conformational sampling.

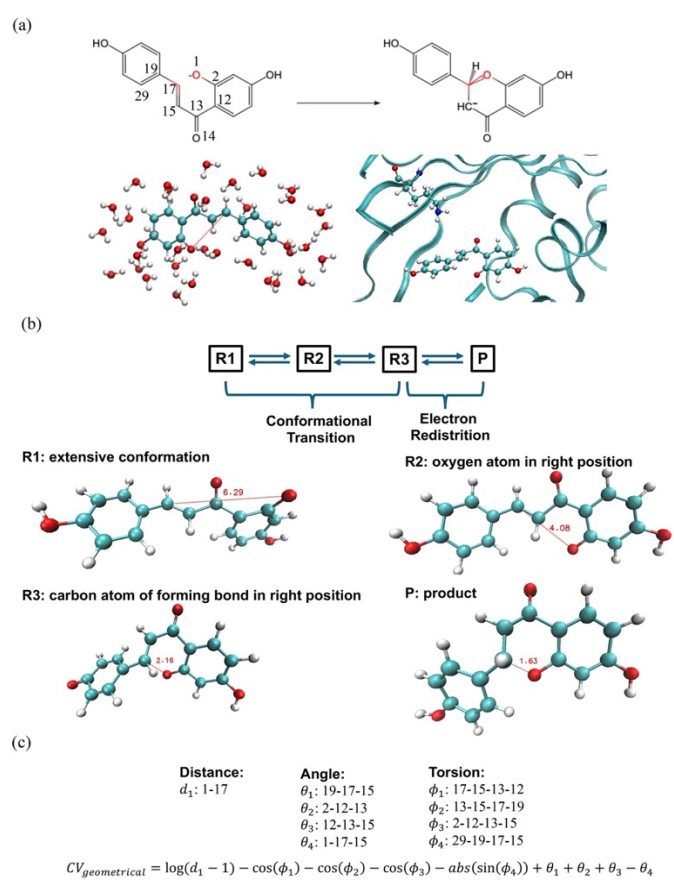

Figure 2. Michael addition of 6'-deoxychalcone in aqueous solution and in the active site of chalcone isomerase (CHI). (a) Reaction schemes in water (left) and in CHI (right). (b) Two coupled stages of the reaction progress: conformational adjustment and electron redistribution. (c) Geometry-based collective variable constructed from chemical intuition, including bond-distance, angular, and dihedral terms.

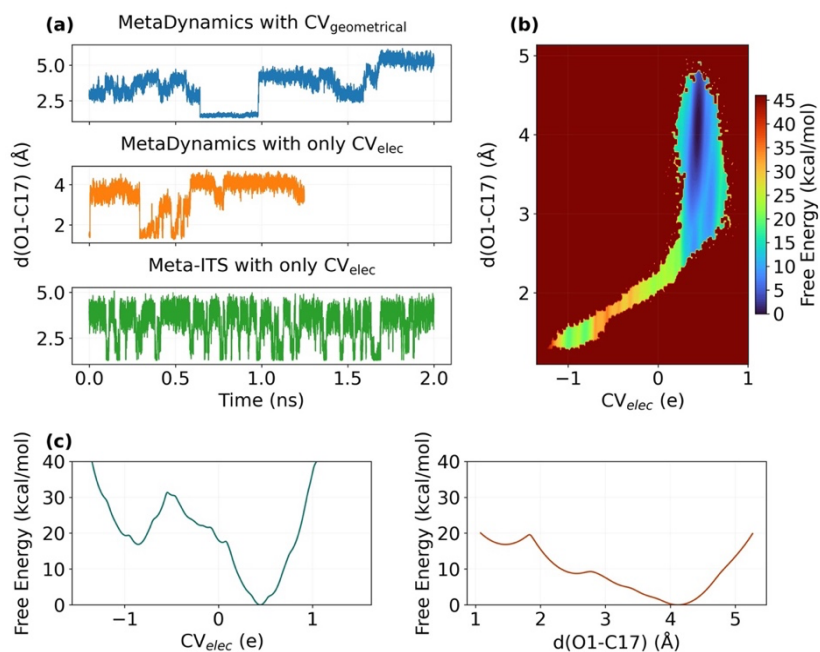

Figure 3. Enhanced sampling of the Michael addition in aqueous solution. (a) Time series of the forming/breaking bond distance O1-C17 obtained from metadynamics simulations with the geometry-based CV (top), the electronic CV alone (middle), and the electronic CV combined with ITS through the Meta-ITS scheme (bottom). (b) Two-dimensional free-energy surface in the O1-C17 distance and electronic CV space obtained from the Meta-ITS simulation. (c) One-dimensional free-energy profiles along the O1-C17 distance and the electronic CV obtained from the Meta-ITS simulation data.

We next applied the electronic CV to the same Michael addition reaction in the more confined enzymatic environment of chalcone isomerase (CHI, Figure 2a) to test how far the charge-space construction can transfer across environments within the same reaction family. Compared with the aqueous system, conformational adjustment in the enzyme is more strongly restricted by the protein environment. In practice, combining the electronic CV with ITS alone was not sufficient to achieve efficient sampling in this enzyme environment, likely because the conformational approach required for reaction is more strongly constrained by the protein matrix. This motivated the inclusion of a more explicit conformational driving component.

We therefore introduced the bonding distance O1-C17 $d(O1 - C17)$,
$$CV_{conf} = d(O1 - C17),$$
into the resulting CV and repeated the metadynamics simulations. With this additional conformational component, reactive events were sampled much more frequently. As in the aqueous system, the electronic CV changed only weakly when the reacting atoms were far apart and became strongly responsive only after the bond-forming atoms approached each other as shown in Figure 4b. This consistent correlation further supports the view that the electronic CV captures the electron-redistribution component of the reaction, whereas the bonding distance mainly reflects the conformational approach. Compared with the aqueous system, the free-energy surfaces in Figure 4b-c and Figure 3b-c show that the enzymatic environment markedly lowers the free-energy cost of electron redistribution, a difference that is most clearly resolved along the electronic CV. Along the bond-distance coordinate, the free-energy difference becomes more consistent with the overall reaction thermodynamics in the enzyme, but the catalytic effect is still much more distinctly revealed along the electronic coordinate. Notably, the same linear coefficients used in the aqueous solution were also applicable in the enzymatic environment, showing that, in this Michael-addition case, the same coefficient set can transfer across environments (See Table S2).

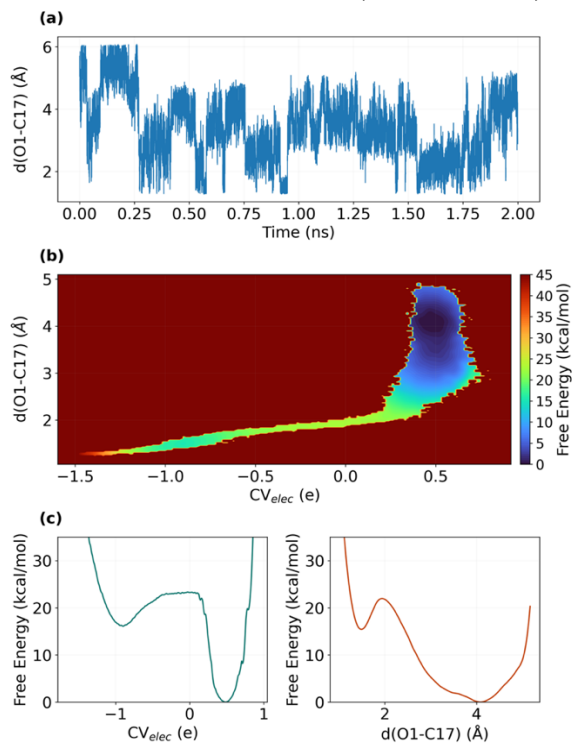

Figure 4. Enhanced sampling of the Michael addition in the CHI active site. (a) Time series of the O1-C17 bond-forming distance obtained from metadynamics simulations using the combined electronic and conformational CV. (b) Two-dimensional free-energy surface in the O1-C17 distance and electronic CV space, showing the conformational and electronic components of the reaction progress in the enzyme environment. (c) One-dimensional free-energy profiles along the O1-C17 distance and the electronic CV.

We next applied the electronic CV to a different reaction type, the Claisen rearrangement of cis-2-vinylcyclopropanecarboxaldehyde (Figure 5a), which interconverts to 2,5-dihydrooxepin, to test whether the same charge-space construction principle remains effective beyond Michael addition. This system was therefore chosen to examine transferability across reaction classes. As in the previous systems, the reaction requires an initial structural adjustment before the subsequent electron redistribution can proceed efficiently.

To describe the conformational component in sampling, we used a simple linear combination of the forming $d(C12 - O15)$ and breaking $d(C3 - C8)$ bond distances,

$$CV_{conf} = d(C3 - C8) - d(C12 - O15),$$
$$s = 0.18 * d(C12 - O15) - 0.82 * d(C3 - C8),$$

while the linear coefficients of the electronic CV were again determined from the sign of the atomic-charge changes (See Table S5). For sampling, we used the simple conformational CV $CV_{conf}$ defined above. For free-energy projection, however, we used a different conformational coordinate, $s$, adopted from previous work[36] as an optimized coordinate for visualization (Figure 5c-d). With this simple combination of conformational and electronic components, reactive transitions were sampled efficiently. Unlike the Michael addition systems, the free-energy surface along the projected conformational coordinate is already highly similar to that along the electronic CV (Figure 5c). This result indicates that, in this Claisen rearrangement, the geometric and electronic degrees of freedom are much more strongly coupled, so that an appropriate combination of bond distances can already capture the electronic component of the reaction rather well. Importantly, the efficient sampling obtained here shows that an explicitly optimized conformational CV is not required for driving the simulation, because even the simple sampling form with coefficients of 1 and -1 is sufficient to work effectively. The successful application of the same charge-based construction principle to this distinct reaction nevertheless supports the transferability of this construction principle across reaction classes.

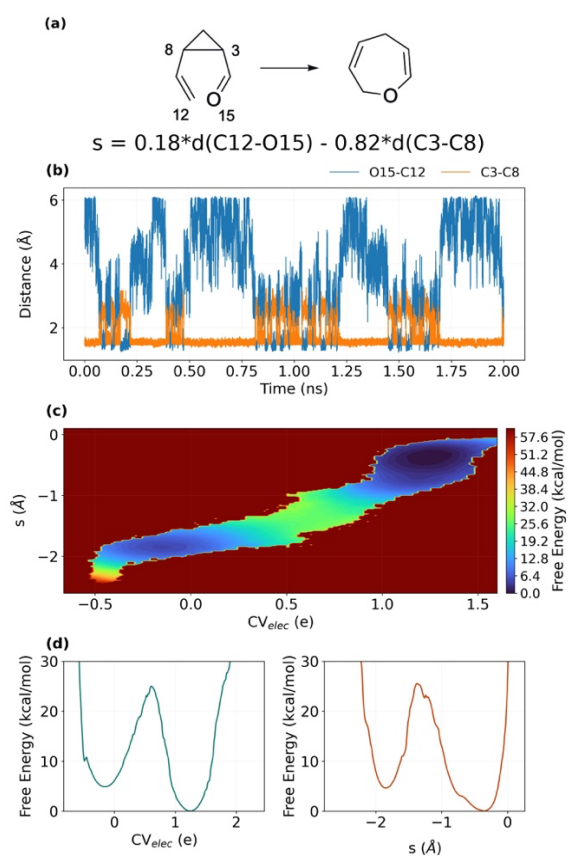

Figure 5. Claisen rearrangement of cis-2-vinylcyclopropanecarboxaldehyde. (a) Reaction scheme. (b) Time series of the forming and breaking bond distances obtained from the metadynamics simulation using the electronic CV together with a simple conformational CV defined as the distance difference with coefficients of 1 and -1. (c) Two-dimensional free-energy surface projected onto the optimized conformational coordinate from previous work, rather than the sampling coordinate, and the electronic CV. (d) One-dimensional free-energy profiles along the projected conformational coordinate and the electronic CV. The projected conformational coordinate and the electronic CV show highly similar free-energy profiles in this system.

We finally considered a more challenging situation in which side reactions compete with the target reaction pathway, in order to extend the framework from reaction driving to reaction selectivity control. Such competing channels are common in enhanced sampling of chemical reactions and are often suppressed by imposing restraining potentials on selected geometrical variables. However, because electron redistribution is typically associated with collective changes in bonds, angles, and dihedrals, constraining undesired pathways in geometric space can become cumbersome and strongly system dependent.

To demonstrate this extension, we applied the framework to the Claisen rearrangement of chorismate to prephenate (Figure 6a). The conformational component was described by the difference between the forming and breaking bond distances,

$$CV_{conf} = d(O17 - C15) - d(C4 - C19),$$

while the linear coefficients of the electronic CV were determined by LDA to distinguish product and reactant states (See Table S7). With this CV, the target reaction could be sampled efficiently; however, side reactions were observed when the accumulated bias became sufficiently large as shown in Figure 6b

To suppress these undesired channels, we defined an additional electronic CV $CV_{elec-res}$ using the same charge-based construction principle, but tailored to distinguish the side-product region from the desired reaction pathway. This additional coordinate was then used to impose a restraining potential during metadynamics simulations. As a result, the side reaction was effectively eliminated while the target reaction remained accessible as shown in Figure 6c.

The two-dimensional free-energy surface in Figure 6d provides further mechanistic insight. At the two sides of the surface, where the bond-forming atoms remain relatively far apart, the free-energy variation is dominated mainly by conformational change. By contrast, in the central region where the reacting atoms approach each other, the free-energy variation is governed predominantly by the electronic coordinate. Moreover, a pronounced bending of the free-energy path is observed near the central region around the horizontal coordinate of 0. and the vertical coordinate of -0.8. This feature suggests that the coupling between conformational and electronic degrees of freedom is more intricate than a simple two-stage picture, with the dominant contribution shifting along the reaction pathway. The one-dimensional free-energy profiles in Figure 6e provide a complementary perspective. Although the barrier height along the conformational coordinate is closer to the value on the full two-dimensional free-energy surface, the trajectory distribution shows that, once the bond-forming atoms have approached each other, the subsequent evolution is dominated by the electronic CV. This contrast suggests that the electronic degree of freedom may play a strong dynamical role, even when its contribution is less obvious from the one-dimensional free-energy barrier alone.

These results show that charge-based electronic CVs can be used not only to promote reactive sampling, but also to control pathway selectivity by electronically discriminating undesired channels from the target transformation. At the same time, this chorismate case study reveals that the interplay between conformational and electronic degrees of freedom can be more complex than a simple sequential picture, underscoring the value of explicitly resolving the electronic component in reaction sampling.

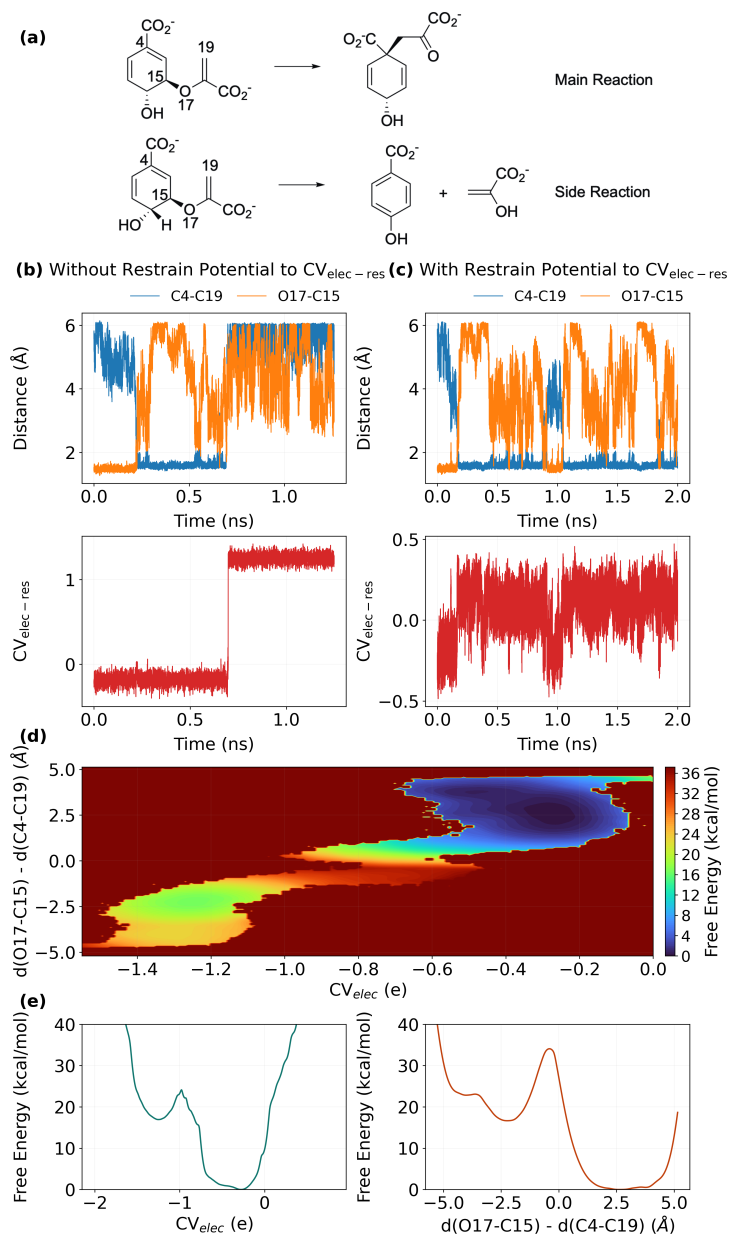

Figure 6. Claisen rearrangement of chorismate and suppression of side reactions by an additional electronic CV. (a) Reaction scheme in aqueous solution. (b) Time series of the forming and breaking bond distances during metadynamics simulations without a restraining potential on the additional electronic CV (top), together with the time series of the additional electronic CV (bottom). (c) Time series of the forming and breaking bond distances during metadynamics simulations with a restraining potential on the additional electronic CV (top), together with the time series of the additional electronic CV (bottom). (d) Two-dimensional free-energy surface in the conformational and electronic CV space. (e) One-dimensional free-energy profiles along the conformational and electronic coordinates.

## 5. Conclusions

In this work, we introduced a charge-based electronic collective variable defined in charge space for chemical reaction sampling. Rather than focusing only on how CVs should be optimized, the present work addresses the more fundamental question of how reaction progress should be represented and in which descriptor space a generally applicable reaction CV should be constructed. By construction, this electronic CV captures the electron-redistribution component of reaction progress and provides a physically motivated alternative to handcrafted geometric descriptors. Across the reactions and environments examined here, the electronic CV can be formulated in a common charge-space form as a linear combination of atomic charges, with the corresponding coefficients assigned in a simple manner from charge differences between relevant states.

A central conclusion of this work is that reaction progress generally involves two coupled components: conformational adjustment and electron redistribution. Accordingly, effective reaction sampling generally requires both a conformational component that brings the system into reactive configurations and an electronic component that captures the subsequent redistribution of electrons. In this sense, the charge-based electronic CV is most useful not as a universal reaction coordinate for all rare events, but as a common and physically grounded descriptor for the electronic component of chemical reaction progress.

In addition to promoting reactive sampling, the same framework can be extended to restrain undesired pathways through additional electronic CVs. This capability demonstrates that charge-based electronic CVs can be used not only to accelerate target reactions, but also to control pathway selectivity by electronically discriminating competing channels. Overall, these results support charge-space electronic CVs as a physically motivated framework for chemical reaction sampling that reduces reliance on handcrafted geometric design while preserving a common representation of the electronic component across the tested systems. In future work, it will be of interest to explore the applicability of this strategy to broader reaction classes and to investigate other electronic descriptors that may further enrich CV design.

## ASSOCIATED CONTENT

## SUPPORTING INFORMATION

The following files are available free of charge.

Figure S1-S6 (PDF)

## AUTHOR INFORMATION

### Corresponding Author

Yi Isaac Yang − Institute of Systems and Physical Biology, Shenzhen Bay Laboratory, Shenzhen 518107, China; orcid.org/0000-0002-5599-0975; Email: yangyi@szbl.ac.cn.

**Authors**

YaoKun Lei − Institute of Systems and Physical Biology, Shenzhen Bay Laboratory, Shenzhen 518107, China; orcid.org/0000-0002-5599-0975; Email: leiyaokun@szbl.ac.cn.

**ACKNOWLEDGMENTS**

This research was supported by National Natural Science Foundation of China (22503057 to Y.K.L, to Y.I.Y). The computational resources were supported by ShenZhen Bay Laboratory Supercomputing Center.

**Notes**

The authors have no conflicts to disclose.

**References**
(1) Unke, O. T.; Chmiela, S.; Sauceda, H. E.; Gastegger, M.; Poltavsky, I.; Schütt, K. T.; Tkatchenko, A.; Müller, K.-R. Machine Learning Force Fields. *Chemical Reviews* **2021**. DOI: 10.1021/acs.chemrev.0c01111.
(2) Behler, J. Four Generations of High-Dimensional Neural Network Potentials. *Chemical Reviews* **2021**. DOI: 10.1021/acs.chemrev.0c00868.
(3) Ceriotti, M.; Clementi, C.; Anatole von Lilienfeld, O. Introduction: Machine Learning at the Atomic Scale. *Chemical Reviews* **2021**. DOI: 10.1021/acs.chemrev.1c00598.
(4) Huang, B.; von Lilienfeld, O. A. Ab Initio Machine Learning in Chemical Compound Space. *Chemical Reviews* **2021**. DOI: 10.1021/acs.chemrev.0c01303.
(5) Nandy, A.; Duan, C.; Taylor, M. G.; Liu, F.; Steeves, A. H.; Kulik, H. J. Computational Discovery of Transition-metal Complexes: From High-throughput Screening to Machine Learning. *Chemical Reviews* **2021**. DOI: 10.1021/acs.chemrev.1c00347.
(6) Hénin, J.; Lelièvre, T.; Shirts, M. R.; Valsson, O.; Delemotte, L. Enhanced Sampling Methods for Molecular Dynamics Simulations [Article v1.0]. *Living Journal of Computational Molecular Science* **2022**. DOI: 10.33011/livecoms.4.1.1583.
(7) Valsson, O.; Tiwary, P.; Parrinello, M. Enhancing Important Fluctuations: Rare Events and Metadynamics from a Conceptual Viewpoint. *Annual Review of Physical Chemistry* **2016**. DOI: 10.1146/annurev-physchem-040215-112229.
(8) Yang, Y. I.; Shao, Q.; Zhang, J.; Yang, L.; Gao, Y. Q. Enhanced sampling in molecular dynamics. *The Journal of Chemical Physics* **2019**. DOI: 10.1063/1.5109531.

(9) Fiorin, G.; Klein, M. L.; Hénin, J. Using collective variables to drive molecular dynamics simulations. *Molecular Physics* **2013**. DOI: 10.1080/00268976.2013.813594.
(10) Sidky, H.; Chen, W.; Ferguson, A. L. Machine learning for collective variable discovery and enhanced sampling in biomolecular simulation. *Molecular Physics* **2020**. DOI: 10.1080/00268976.2020.1737742.
(11) Chen, M. Collective variable-based enhanced sampling and machine learning. *The European Physical Journal B* **2021**. DOI: 10.1140/epjb/s10051-021-00220-w.
(12) Mehdi, S.; Smith, Z.; Herron, L.; Zou, Z.; Tiwary, P. Enhanced Sampling with Machine Learning. *Annual Review of Physical Chemistry* **2024**. DOI: 10.1146/annurev-physchem-083122-125941.
(13) Best, R. B.; Hummer, G. Reaction coordinates and rates from transition paths. *Proceedings of the National Academy of Sciences* **2005**. DOI: 10.1073/pnas.0408098102.
(14) Jung, H.; Covino, R.; Arjun, A.; Leitold, C.; Dellago, C.; Bolhuis, P. G.; Hummer, G. Machine-guided path sampling to discover mechanisms of molecular self-organization. *Nature Computational Science* **2023**. DOI: 10.1038/s43588-023-00428-z.
(15) Chen, H.; Roux, B.; Chipot, C. Discovering Reaction Pathways, Slow Variables, and Committor Probabilities with Machine Learning. *Journal of Chemical Theory and Computation* **2023**. DOI: 10.1021/acs.jctc.3c00028.
(16) Megías, A.; Contreras Arredondo, S.; Chen, C. G.; Tang, C.; Roux, B.; Chipot, C. Iterative variational learning of committor-consistent transition pathways using artificial neural networks. *Nature Computational Science* **2025**. DOI: 10.1038/s43588-025-00828-3.
(17) Li, Q.; Lin, B.; Ren, W. Computing committor functions for the study of rare events using deep learning. *The Journal of Chemical Physics* **2019**. DOI: 10.1063/1.5110439.
(18) Kang, P.; Trizio, E.; Parrinello, M. Computing the committor with the committor to study the transition state ensemble. *Nature Computational Science* **2024**. DOI: 10.1038/s43588-024-00645-0.
(19) Zhang, J.; Lei, Y.-K.; Zhang, Z.; Han, X.; Li, M.; Yang, L.; Yang, Y. I.; Gao, Y. Q. Deep reinforcement learning of transition states. *Physical Chemistry Chemical Physics* **2021**. DOI: 10.1039/d0cp06184k.
(20) Lei, Y.-K.; Zhang, Z.; Han, X.; Yang, Y. I.; Zhang, J.; Gao, Y. Q. Locating Transition Zone in Phase Space. *Journal of Chemical Theory and Computation* **2022**. DOI: 10.1021/acs.jctc.2c00385.
(21) Klus, S.; Nüske, F.; Koltai, P.; Wu, H.; Kevrekidis, I.; Schütte, C.; Noé, F. Data-Driven Model Reduction and Transfer Operator Approximation. *Journal of Nonlinear Science* **2018**. DOI: 10.1007/s00332-017-9437-7.
(22) Wu, H.; Nüske, F.; Paul, F.; Klus, S.; Koltai, P.; Noé, F. Variational Koopman models: Slow collective variables and molecular kinetics from short off-equilibrium simulations. *The Journal of Chemical Physics* **2017**. DOI: 10.1063/1.4979344.

(23) Schütte, C.; Klus, S.; Hartmann, C. Overcoming the timescale barrier in molecular dynamics: Transfer operators, variational principles and machine learning. *Acta Numerica* **2023**. DOI: 10.1017/s0962492923000016.
(24) Gaus, M.; Cui, Q.; Elstner, M. DFTB3: Extension of the Self-Consistent-Charge Density-Functional Tight-Binding Method (SCC-DFTB). *Journal of Chemical Theory and Computation* **2011**. DOI: 10.1021/ct100684s.
(25) Yang, L.; Gao, Y. Q. A selective integrated tempering method. *The Journal of Chemical Physics* **2009**. DOI: 10.1063/1.3266563.
(26) Lei, Y.-K.; Yagi, K.; Sugita, Y. Efficient Training of Neural Network Potentials for Chemical and Enzymatic Reactions by Continual Learning. *Journal of Chemical Theory and Computation* **2025**. DOI: 10.1021/acs.jctc.4c01393.
(27) Lei, Y.-K.; Yagi, K.; Sugita, Y. Learning QM/MM potential using equivariant multiscale model. *The Journal of Chemical Physics* **2024**. DOI: 10.1063/5.0205123.
(28) Jo, S.; Kim, T.; Iyer, V. G.; Im, W. CHARMM-GUI: A Web-Based Graphical User Interface for CHARMM. *Journal of Computational Chemistry* **2008**. DOI: 10.1002/jcc.20945.
(29) Suh, D.; Thodika, A. R. A.; Kim, S.; Nam, K.; Im, W. CHARMM-GUI QM/MM Interfacer for a Quantum Mechanical and Molecular Mechanical (QM/MM) Simulation Setup: 1. Semiempirical Methods. *Journal of Chemical Theory and Computation* **2024**. DOI: 10.1021/acs.jctc.4c00439.
(30) Hourahine, B.; Aradi, B.; Blum, V.; Bonafé, F.; Buccheri, A.; Camacho, C.; Cevallos, C.; Deshaye, M. Y.; Dumitrică, T.; Dominguez, A.; et al. DFTB+, a software package for efficient approximate density functional theory based atomistic simulations. *The Journal of Chemical Physics* **2020**. DOI: 10.1063/1.5143190.
(31) Hourahine, B.; Berdakin, M.; Bich, J. A.; Bosia, F.; Chou, C.-P.; Cusati, T.; Dall'Olio, S.; Dominguez, A.; Dupuy, R.; Elstner, M.; et al. Recent Developments in DFTB+, a Software Package for Efficient Atomistic Quantum Mechanical Simulations. *The Journal of Physical Chemistry A* **2025**. DOI: 10.1021/acs.jpca.5c01146.
(32) Huang, Y.-P.; Xia, Y.; Yang, L.; Wei, J.; Yang, Y. I.; Gao, Y. Q. SPONGE: A GPU-Accelerated Molecular Dynamics Package with Enhanced Sampling and AI-Driven Algorithms. *Chinese Journal of Chemistry* **2022**. DOI: 10.1002/cjoc.202100456.
(33) Gaus, M.; Goez, A.; Elstner, M. Parametrization and Benchmark of DFTB3 for Organic Molecules. *Journal of Chemical Theory and Computation* **2013**. DOI: 10.1021/ct300849w.
(34) Laio, A.; Parrinello, M. Escaping free-energy minima. *Proceedings of the National Academy of Sciences* **2002**. DOI: 10.1073/pnas.202427399.
(35) Yang, Y. I.; Niu, H.; Parrinello, M. Combining Metadynamics and Integrated Tempering Sampling. *The Journal of Physical Chemistry Letters* **2018**. DOI: 10.1021/acs.jpclett.8b03005.
(36) Zhang, J.; Zhang, Z.; Yang, Y. I.; Liu, S.; Yang, L.; Gao, Y. Q. Rich Dynamics Underlying Solution Reactions Revealed by Sampling and Data Mining of Reactive Trajectories. *ACS Central Science* **2017**. DOI: 10.1021/acscentsci.7b00037.